\documentstyle[12pt,epsf]{article}

\newcommand{\gsim}{\lower.7ex\hbox{$\;\stackrel{\textstyle>}{\sim}\;$}}
\newcommand{\lsim}{\lower.7ex\hbox{$\;\stackrel{\textstyle<}{\sim}\;$}}

\def\beq{\begin{equation}}
\def\eeq{\end{equation}}
\def\bea{\begin{eqnarray}}
\def\eea{\end{eqnarray}}
\def\bq{\begin{quote}}
\def\eq{\end{quote}}

\def\bq{\begin{quote}}
\def\eq{\end{quote}}

\begin{document}

\baselineskip 24pt
\newcommand{\sheptitle}
{Naturalness Implications of LEP Results}

\newcommand{\shepauthor}
{G. L. Kane$^1$ and S. F. King$^2$ }

\newcommand{\shepaddress}
{$^1$Randall Physics Laboratory, University of Michigan,
Ann Arbor, MI 48109-1120\\
$^2$Department of Physics and Astronomy,
University of Southampton, Southampton, SO17 1BJ, U.K.}

\newcommand{\shepabstract} {We analyse the fine-tuning constraints
arising from absence of superpartners at LEP,
without strong universality assumptions. We show that such constraints
do not imply that charginos or neutralinos should have been seen at LEP,
contrary to the usual arguments.  They do however imply relatively light
gluinos $\left( m_{\tilde g} \lsim 350 \rm{GeV}\right)$ and/or a relation
between the soft-breaking $SU(3)$ gaugino mass and Higgs soft mass
$m_{H_U}$.
The LEP limit on the Higgs mass is significant, 
especially at low $\tan \beta$, and we investigate
to what extent this provides evidence for both a lighter gluino and
correlations between soft masses.}

\begin{titlepage}
\begin{flushright}
hep-ph/9810374\\
SHEP-98/13
\end{flushright}

\begin{center}
{\large{\bf \sheptitle}}
\bigskip \\ \shepauthor \\ \mbox{} \\ {\it \shepaddress} \\ \vspace{.5in}
{\bf Abstract} \bigskip \end{center} \setcounter{page}{0}
\shepabstract
\begin{flushleft}
\today
\end{flushleft}
\end{titlepage}

\section{Introduction}

There are several circumstantial pieces of evidence for supersymmetry
(SUSY) not least of which are the measured magnitudes of the gauge
coupling strengths at low energy which hint at a high energy unification
for a light SUSY spectrum.  This fits in
well with a prime motivation for SUSY which is to protect the Higgs mass
parameter from large radiative corrections, which again requires a light
SUSY spectrum. Taken together these self-reinforcing pieces of
circumstantial evidence provide rather convincing phenomenological
support for what would in any case be an elegant extension of the
standard model.  However although the unification and fine-tuning
arguments look quite convincing there are two areas of concern coming
from recent LEP measurements. Our philosophy is that such areas of
concern should be regarded as opportunities for new insights.

One such opportunity
is the discrepancy between the
world average experimental measurement of the strong coupling
$\alpha_3(M_Z)=0.119(4)$ and the value predicted from SUSY GUTs
$\alpha_3(M_Z)=0.13(1)$, an effect we think is significant even though
the theoretical uncertainty is hard to quantify\cite{Lang}.  This
difference may be telling us about aspects of the spectrum or of Planck
scale physics.  A second opportunity is due to the recent LEP
limits on the mass of SUSY and Higgs particles and the implications this
has for fine-tuning.  In particular the absence of both charginos and
Higgs at the highest energy LEP runs has been argued to significantly
increase the fine-tuning ``price'' relative to what it was previously
\cite{price}.  Earlier analyses which developed these fine-tuning
arguments had raised expectations that SUSY particles should be found at
LEP \cite{earlier}.

Here we take a constructive point of view that these mild quantitative
difficulties referred to above are in fact useful pointers which are
telling us something important about the SUSY spectrum.  A common
assumption, motivated by minimal supergravity (SUGRA), is that at high
energies the SUSY spectrum is described by a universal soft scalar mass,
$m_0$, and universal gaugino mass, $M_{1/2}$, which together with the
universal trilinear parameter $A_0$ and the bilinear parameter $B_0$ and
Higgs superpotential mass parameter $\mu_0$ form the five high energy
input parameters of the constrained minimal supersymmetric standard
model (CMSSM) \cite{cmssm}.  Many of the unification and fine-tuning
analyses presented in the literature are based completely or partially
on the CMSSM, and of those analyses that do relax universality nearly
all do so in the scalar sector while maintaining gaugino mass
universality. From these assumptions, and the absence of superpartners at
LEP, it is concluded by some authors that we should be nervous about the
validity of the general SUSY framework.  An exception is the
unification analysis of Roszkowski and Shifman \cite{RS} which showed
that by relaxing the assumption of gaugino mass unification, and
allowing the gluino mass to be smaller than the wino mass at the GUT
scale, $M_3(0)< M_2(0)$, the effect of low energy thresholds due to a
lighter gluino and heavier wino is to push down the SUSY prediction of
$\alpha_3(M_Z)$ towards the experimental value.  
We will see below that the natural
solution of the fine-tuning problem also reduces $\alpha_3(M_Z)$, perhaps
increasing our confidence in the relevence of our results.

Another exception is the fine-tuning analysis of Wright \cite{Wright}
which we were unaware of in a previous version of this paper.
Wright also relaxes universality of the gaugino masses, 
and sets fine-tuning limits on the masses of superpartners.
He concludes that the most problematic constraint arises from
the gluino mass which is bounded to be less that 260 GeV 
corresponding to a fine-tuning limit of 10\%.
He also sets a fine-tuning limit on the $\mu$ parameter of 140 GeV.
In hindsight the present paper serves to reaffirm Wright's conclusions
by including some details left out of the calculation of Wright
(such as the running of the top quark Yukawa coupling)
and also gives a qualitative analytic understanding of why the
gluino mass plays such an important role in fine-tuning.
We shall also address the role of the Higgs mass in fine-tuning,
which was not covered by Wright.
In addition we shall propose a genuinely new idea 
(as far as we are aware) which 
we call the supernatural solution to the fine-tuning opportunity.

In this paper, then, we investigate fine-tuning without the
gaugino mass universality assumption
\footnote{ We emphasise that gauge coupling unification is {\it a priori}
independent of the question of gaugino mass unification.}. 
We focus on this because the usual results most strongly
constrain charginos and neutralinos.  
We explore two ways to obtain small fine-tuning: 
\begin{enumerate}
\item We show that having a gluino mass smaller than its
universality value reduces the
fine-tuning dramatically;
\item  We suggest new theoretical correlations or simply relations
between input parameters in order to reduce fine-tuning, such as
a correlation or relation between $m_{H_U}$, the soft-breaking scalar
mass that triggers the Higgs mechanism,
and $M_3$, the $SU(3)$ gaugino mass that is proportional to the gluino
mass, or between $\mu$ and the gluino mass (if $\mu$ is large).
\footnote{Chankowski et al \cite{price} have also
recently examined reducing fine-tuning by correlations among parameters,
and the possible connection to string theory.  Since they maintain
gaugino mass
universality their conclusions are  different from ours.}
\end{enumerate}

We emphasise the distinction between a correlation and a relation
between two input parameters. By a correlation we mean an exact
theoretical relation such that the two parameters should no longer
be regarded as independent, but instead should be regarded as a 
single input parameter. A relation on the other hand is simply
a statement of the relative magnitude of one parameter relative
to another. Clearly correlations have the potential to completely
or partially alleviate fine-tuning if their correlation
involves a partial cancellation of two large numbers to obtain a 
numerically small physical quantity,
although the origin of such a desirable correlation may remain
obscure. By contrast simple relations between parameters
may lead to modest reductions in fine-tuning by simply moving 
to a different part of non-universal parameter space which is 
more favourable from the point of reducing the total
(or global) amount of fine-tuning. We shall discuss 
correlations in the text and explore relations numerically.

Which of 1 or 2 holds, or whether both do, is fully testable
experimentally.  For example the first
implies a physical $\tilde g$ mass less than about
350 GeV, possibly even around 250 GeV. For $\tan \beta \sim 1$
correlations are the only
option, because the absence of a Higgs boson at LEP does not allow the
gluino mass to be light enough to significantly reduce the fine-tuning
(see below).

We emphasise that in the first case above
we do not envisage a {\em very} light gluino such that
it becomes the lightest supersymmetric particle (LSP)
\cite{lightgluino}, but merely that it is somewhat lighter than the
universal gaugino mass prediction, by an amount which we shall quantify
in due course but which typically may be around 1/2 of its universal value.
Also even though we abandon minimal SUGRA we shall stay
within the general framework of the SUGRA mechanism of SUSY breaking for
the present discussion, though many of our conclusions will apply also
to gauge mediated SUSY breaking scenarios. We should also point out that
the abandonment of universality fits in well with current ideas of
string and M-theory in whose framework the minimal SUGRA model, which
provided the theoretical impetus for the CMSSM, plays no special role.
On the other hand the absence of flavour-changing neutral currents
(FCNC's) seems to imply that there must be a high degree of universality
at least in flavour space.  These arguments lead one to expect that the
high energy soft SUSY breaking parameters could well contain as much
non-universality as can be tolerated consistent with the absence of
FCNC's. 

Following the above philosophy leads to the recently proposed so called minimal
reasonable model (MRM) \cite{MRM}. 
From the point of view of
our solution to the fine-tuning problem the phases \cite{MRM, phases}
play no special
role though they will certainly have quantitative effects if they are included,
but we set them to zero for our qualitative study.
We shall restrict ourselves to low and intermediate $\tan \beta$
for this analysis.
With these additional assumptions the MRM depends only on the
top quark Yukawa coupling $h_t(0)$ plus the
following 13 real parameters $a$ at the unification scale $M_{U}$ where:
\bea
a\in \{m_Q^2(0),m_U^2(0),m_D^2(0),m_L^2(0),m_E^2(0),m_{H_D}^2(0),m_{H_U}^2(0),
\nonumber \\
M_1(0),M_2(0),M_3(0),A_t(0),B(0),\mu(0)\}
\label{highparameters}
\eea 
The notation is such that $t=\ln (M_U^2/Q^2)$ where $Q$ is the
$\bar{MS}$ energy scale, so that $t=0$ corresponds to $Q=M_U$.  We
do not have to choose between GUT and string unification. The version of
MRM we use here includes an independent soft mass parameter for every
chiral multiplet of the MSSM $Q^i,U^i,D^i,L^i,E^i,H_D,H_U$ in a common
notation corresponding to quark doublets, up-type quark singlets,
down-type quark singlets, lepton doublets, charged lepton singlets,
Higgs doublet coupling to down-type quarks and charged leptons, Higgs
doublet coupling to up-type quarks, respectively. At high energies the
soft mass matrices in family space are assumed to be diagonal and
family-independent (given by the 5 squark and slepton mass parameters
in Eq.\ref{highparameters} multiplying unit matrices) as motivated
by the phenomenological requirement of acceptable FCNC's.  We are also
implicitly assuming that the only soft trilinear parameter of importance
is the one corresponding to the top Yukawa coupling, which again seems
reasonable. We should also point out that
the above parameter set in Eq.\ref{highparameters} may be
more than just an arbitrary choice. For example if the gaugino masses
are large compared to the scalar masses 
at the string scale then the parameter set
in Eq.\ref{highparameters} 
(with additional relations between the soft masses)
may be reproduced at a slightly lower scale
as an infra-red fixed point in a class of theories \cite{GG}

One loop semi-analytic solutions to the renormalisation group equations
corresponding to the above parameter set in Eq.\ref{highparameters}
have been presented in
ref.\cite{bottomup} whose sign conventions for $A_t, \mu, B$ we
adopt. The solutions represent an extension of those in ref.\cite{Ibanez}.
The existence of analytic solutions in which the low energy diagonal masses
\bea
\{m_{Qi}^2(t),m_{Ui}^2(t),m_{Di}^2(t),m_{Li}^2(t),m_{Ei}^2(t)
,m_{H_D}^2(t),m_{H_U}^2(t),
\nonumber \\
M_1(t),M_2(t),M_3(t),A_t(t),B(t),\mu(t)\}
\label{lowparameters}
\eea
may be expressed in terms of the high energy parameters in
Eq.\ref{highparameters} is important because it enables
fine-tuning (and the conditions for its absence) 
to be understood at a qualitative
level. 

\section{Analysis}

Let us begin our discussion of fine-tuning by recalling a few basic
features of electroweak symmetry breaking in the MSSM, at the RG
improved tree-level. The potential is: 
\beq
V_0=m_1^2v_1^2+m_2^2v_2^2-2m_3^2v_1v_2+\frac{1}{8}(g'^2+g_2^2)(v_1^2-v_2^2)^2 
\eeq
where $v_1,v_2$ are the (assumed) neutral Higgs vacuum expectation
values (VEVs), $g_2,g'$ are the $SU(2)_L\times U(1)_Y$ gauge couplings
and the mass parameters $m_1^2,m_2^2,m_3^2$ are evaluated at low energy
and are given by \beq m_1^2(t)=m_{H_D}^2(t)+\mu^2(t), \ \
m_2^2(t)=m_{H_U}^2(t)+\mu^2(t), \ \ m_3^2(t)=B(t)\mu (t) \eeq The
conditions for successful electroweak symmetry breaking are \bea
m_1^2(t)m_2^2(t)- m_3^4(t)<0 \\ m_1^2(t)+m_2^2(t)- 2m_3^2(t)>0 \eea
where the first condition makes the symmetric case unstable, and the
second condition ensures that the potential is bounded.  These
conditions are achieved in practice by virtue of the large top Yukawa
coupling which drives $m_2^2(t)$ small and often negative.  Assuming
these conditions are met the minimisation conditions are expressable as:
\bea 
\frac{M_Z^2}{2} & = & -\mu^2(t) + \left(
\frac{m_{H_D}^2(t)-m_{H_U}^2(t)\tan^2 \beta}{\tan^2 \beta -1} \right) 
\label{7} \\
\sin 2\beta & = & \frac{2m_3^2(t)}{m_1^2(t)+m_2^2(t)} 
\eea 
where \beq \tan
\beta =v_2/v_1, \ \ M_Z^2=\frac{1}{2} (g'^2+g_2^2)(v_1^2+v_2^2) 
\eeq 
The $U(1)_Y$ coupling normalised appropriately for unification
is given by $g_1^2=(5/3)g'^2$.

The principle of the electroweak symmetry breaking mechanism is that the
high energy parameters in Eq.\ref{highparameters} are fixed
(presumably by some SUSY breaking mechanism in string theory). 
Then they run down to low energies and induce
electroweak symmetry breaking along the above lines, with the Z mass and
$\tan \beta$ predicted from a given set of input parameters.
\footnote{The above discussion has
neglected the crucial one-loop Coleman-Weinberg corrections but the
basic principles remain the same when these are included.
In fact it is well appreciated in refs.\cite{price}, \cite{earlier}
that the effect of such corrections actually helps to
stabilise the electroweak scale and reduce fine-tuning,
so any discussion which neglects them represents a sort
of worst-case situation.}

The basic question of fine-tuning is one of the sensitivity
of the electroweak scale, in this case expressed as the Z mass
squared, to small variations in the input parameters which are
perturbed around a particular physical solution corresponding
to an acceptable Z mass. At a very basic level one may see the
fine-tuning explicitly by expanding the formula for the Z mass
squared Eq.\ref{7} in terms of the low energy masses in 
Eq.\ref{lowparameters} which in turn
are expressed as a function of high energy input parameters
in Eq.\ref{highparameters} via the analytic solutions.
For example for $\tan \beta =2.5$ we find
\bea
\frac{M_Z^2}{2} =  
& & \mbox{} - .87\,\mu^2(0) +
3.6\,{M_3^2(0)}- .12\,{ M_2^2(0)} + .007\,{M_1^2(0)}  \nonumber \\
& & \mbox{} - .71\, {m_{H_U}^2(0)} + .19\,{ m_{H_D}^2(0)}
+ .48\,({ m_Q^2(0)} + \,{m_U^2(0)}) 
\nonumber \\
 & & \mbox{}  - .34\,{A_t(0)}\,{M_3(0)} - .07\,{ A_t(0)}\,{M_2(0)} 
- .01\,{A_t(0)}\,{M_1(0)}
+ .09\,{ A_t^2(0)}\nonumber \\
 & & \mbox{} + .25\,{M_2(0)}\,{M_3(0)}+ .03\,{M_1(0)}\,{M_3(0)} 
+ .007\,{M_1(0)}\,{M_2(0)}
\label{MZ}
\eea
This is an important equation to examine.  First, notice that the
coefficients of terms involving
$M_1(0)$ and $M_2(0)$, the soft mass parameters that determine
the chargino and neutralino masses, are very small 
compared to those involving the high energy gluino mass $M_3(0)$.
For $M_3(0)\gg M_Z$ it is necessary to tune $\mu^2 (0)$ for a
given $\tan \beta$ (which in turn is fixed by adjusting $B$)
so that the large terms are cancelled and
the correct Z mass is achieved. This is the fine-tuning problem.
The solution to the fine-tuning problem is clearly to reduce
$M_3(0)$ as much as possible since it has by far the largest coefficient.
By contrast $M_1(0)$ and $M_2(0)$
may be increased almost arbitrarily without affecting fine-tuning.
For $\tan \beta =1.5,2.5,10$
the coefficient of the $M_3^2(0)$ term is $6.1,3.6,2.8$ which shows
that the fine-tuning problem is worse for low $\tan \beta$.
Thus we conclude that the fine-tuning analysis mainly limits
the gluino mass and only constrains the chargino and
neutralino masses very weakly.

One can see from Eq.\ref{MZ} that there is a second way to overcome
the apparent fine-tuning problem, namely to invoke theoretical
correlations between $M_3(0)$ and any parameter which enters with
a negative sign, such as $\mu(0), m_{H_U}^2(0), A_t(0), M_2(0)$.
The idea of a correlation between $\mu(0)$ and the universal gaugino mass
$M_{1/2}$ is well known (see for example Chankowski et al \cite{price}).
From our point of view this
would become a correlation between $\mu(0)$ and $M_3(0)$,
and it could only be relevant for large $\mu (0)$.
We can identify other perhaps more plausible correlations.
For example, suppose that the fundamental theory after supersymmetry
breaking led to $m_{H_U}(0)\approx 2M_3(0)$.  Then the
combination of the $m^2_{H_U}(0)$ and $M_3^2(0)$ terms has a coefficient
less than or of order unity, as do all other terms, and there is no
fine-tuning problem. Presumably other terms would enter into the true
relation, but the essential feature is that between $m_{H_U}(0)$ and
$M_3(0)$. We regard such correlations between soft masses to be more 
likely than a correlation involving $\mu$ which is not a soft mass.

If relatively light gluinos are found
experimentally (say 200 $\lsim m_{\tilde g} \lsim 350\> {\rm GeV})$ then
apparently the fine-tuning
problem is solved by small $M_3(0)$, while if the
gluino mass is larger it implies correlations such as between $M_3(0)$ and 
$m_{H_U}(0)$.  Such a relation would be an important
clue to the form of SUSY
breaking of string theory.  By such reasoning we could
learn about the form of unification or Planck scale physics from
collider data, even before superpartners are found!

The above qualitative 
discussion takes no account of the implicit sensitivity of
the Z mass coming from changes in $\tan \beta$ as a result of
small variations in the high energy inputs.
This is addressed by the master formula of Dimopoulos and Giudice
\cite{earlier} which yields a fine-tuning parameter which corresponds
to the fractional change in the Z mass squared per unit fractional
change in the input parameter,
\beq
\Delta_a=abs \left( \frac{a}{M_Z^2}\frac{\partial M_Z^2}{\partial a}\right)
\eeq
for each input parameter $a$ in Eq.\ref{highparameters}.
\footnote{An alternative definition of fine-tuning replaces $M_Z^2$
by $M_Z$, which leads to fine-tuning parameters only half as
large as our estimates. 
Both definitions are used in the literature \cite{price},
sometimes by the same authors.}
Following the above discussion
it is sufficient for our purposes to calculate three parameters
$\Delta_{\mu(0)}$, $\Delta_{M_3(0)}$ and $\Delta_{m_{H_U}^2(0)}$. 
As with the Z mass we may expand these fine-tuning
parameters in terms of the high energy input parameters, 
and so investigate the source of the fine-tuning.
For $\tan \beta =2.5$ we find
\bea
\Delta_{\mu(0)} & \approx &  
5.1\tilde{\mu}^2 (0)+
2.7\,{\tilde{M}_3^2(0)}- 0.6\,{\tilde{M}_2^2(0)}
- .03\,{\tilde{M}_1^2(0)} 
\nonumber \\
& - & .54\,{\tilde{m}_{H_U}^2(0)} - .91\,{\tilde{m}_{H_D}^2(0)}
+ .36\,({\tilde{m}_{Q}^2(0)}+ \,{\tilde{m}_U^2(0)})
\nonumber \\
& - & .26\,{\tilde{A}_t(0)}\,{\tilde{M}_3(0)}
- .06\,{\tilde{A}_t(0)}\,{\tilde{M}_2(0)}
- .009\,{\tilde{A}_t(0)}\,{\tilde{M}_1(0)}+   .07\,{\tilde{A}_t(0)}^{2}
\nonumber \\
& + & .19\,{\tilde{M}_2(0)}\,{\tilde{M}_3(0)}
+ .03\,{\tilde{M}_1(0)}\,{\tilde{M}_3(0)}
+.005\,{\tilde{M}_1(0)}\,{\tilde{M}_2(0)} 
\label{Deltamu}
\eea
where the tilde denotes that the parameter is scaled by $M_Z$.
It is possible to eliminate $\mu^2(0)$ using Eq.\ref{MZ}, which would
lead to a dominant term $23.7\,{\tilde{M}_3^2(0)}$.
For $\tan \beta =1.5,2.5,10$
the coefficient of the $\tilde{M}_3^2(0)$ term is
$115,24,12$.

As with Eq.\ref{MZ}, Eq.\ref{Deltamu} shows that
there is greatly
reduced sensitivity to $M^2_1 (0)$ or $M^2_2 (0)$ compared to
$M_3^2(0)$. 
The fine-tuning from Eq.\ref{Deltamu}
is in fact much worse than anticipated from Eq.\ref{MZ} due to 
the implicit $\tan \beta$ sensitivity.
It is clear that once either 
$\tilde{M}_3(0)$ or $\tilde{\mu}^2(0)$ become
larger than unity then the amount of fine-tuning grows quadratically.
The effect is ameliorated to some extent
by the negative contributions coming from
${\tilde{m}_{H_U}^2(0)}$, ${\tilde{M}_2^2(0)}$ and the terms
proportional to ${\tilde{A}_t(0)}$. But such terms
cannot be made arbitrarily large since then the fine-tuning
parameters associated with them will become important, unless
we postulate some theoretical correlation between them.

Having shown that the gluino mass parameter is largely responsible for
fine-tuning, let us briefly take stock of how this comes
about. Naively, one might think that the physics of electroweak
symmetry breaking (EWSB)
is purely related to physics in the electroweak
sector, and the existence of a large top Yukawa coupling.  As it is
normally portrayed, $m^2_{H_U}$  runs and becomes negative if one simply
includes a large top Yukawa in its renormalisation group equation
(RGE). However QCD plays an important role in EWSB by ensuring that
the squark masses, and especially $m^2_{U_3}$ is not driven more
negative, leading to charge and colour breaking minima being
preferred. The Higgs mass parameter $m^2_{H_U}$ feels QCD effects
via the top Yukawa coupling (which itself depends on QCD effects
for its magnitude) and the gluino mass which thereby enters turns
out to dominate due to the relatively fast variation of the
QCD coupling as compared to the electroweak couplings:
\beq
\left(\frac{\alpha_1(t)}{\alpha_1(0)} \right) \approx \frac{5}{12},\ \
\left(  \frac{\alpha_2(t)}{\alpha_2(0)} \right) \approx \frac{10}{12},\ \
\left(  \frac{\alpha_3(t)}{\alpha_3(0)} \right) \approx \frac{32}{12},
\eeq
where $t=\ln (M_U^2/M_Z^2)\approx 66$.

\section{Higgs Mass Constraints}

Having argued that the absence of charginos and neutralinos
at LEP is only weakly related to 
fine-tuning issues, now let us turn to the question of the lightest CP
even Higgs boson (henceforth simply called the Higgs).  
At tree-level the Higgs mass is given by
$m_h^0\approx M_Z|\cos 2 \beta |$ with $m_h^0=35,66,89$ GeV for
$\tan \beta =1.5,2.5,10$, to be compared to
the current LEP limit on the standard model
Higgs mass roughly given by about
90 GeV but being reduced to 75 GeV or so (when SUSY effects and phases
are fully included). 

It is well known that radiative corrections
play an important role for the Higgs \cite{higgsrc}.
A simplified
 expression for radiatively corrected Higgs mass includes the terms
\beq m_h^2\approx M_Z^2 \cos^2 2 \beta + (34 \ GeV)^2 \ln
\left( \frac{m_{\tilde{t}_1}^2 m_{\tilde{t}_2}^2}{m_t^4}\right) + \ldots 
\label{rc}
\eeq
where the ellipsis represents more complicated terms. It is
clear that the radiative corrections depend logarithmically on the
determinant of the matrix of stop masses, at least for the second
term, and in a more complicated way for the remaining terms.
We can expand the determinant of the
stop matrix, which involves the off-diagonal element
$m_t(A_t-\mu \cot \beta )$,
\footnote{Phases can have a significant effect here because they 
can change the coherence of $A_t$ and $\mu$ \cite{MRM}.}
as a function of the high energy input parameters,
and for $\tan \beta =2.5$ we find a lengthy expression
whose leading terms are,
\bea
{m_{\tilde{t}_1}^2m_{\tilde{t}_2}^2} & \approx &
(170 \ GeV)^4 + 26\,{ M_3^4(0)}+ 1.5\,{ A_t(0)}\,{ M_3^3(0)} \nonumber
\\
& & \mbox{} + \left(2.5\,{ m_Q^2(0)} + 3.5\,{ m_U^2(0)} -2.1\,{m_{H_U}^2(0)}
\right) \,{ M_3^2(0)} \nonumber \\
 & & \mbox{} + (134 \ GeV)^2\,{m_Q^2(0)} + (130 \ GeV)^2\,{m_U^2(0)}
- (108 \ GeV)^2\,{m_{H_U}^2(0)} \nonumber \\
& & \mbox{} + (430 \ GeV)^2\,{ M_3^2(0)}
-(180 \ GeV)^2M_2(0)M_3(0) \nonumber \\
 & & \mbox{}  + (180 \ GeV)^2\,{A_t(0)}\,{M_3(0)} - (59 \ GeV)^2\,{A_t^2(0)}
- (65 \ GeV)^2\,\mu^2 (0) \nonumber \\
 & & \mbox{} + (67 \ GeV)^2\,{A_t(0)}\,\mu (0)
- (210 \ GeV)^2\,{M_3(0)}\,\mu (0) 
\label{det}
\eea
Taken together Eqs.\ref{rc} and \ref{det} show that any
shortfall in the tree-level contribution to the Higgs mass
must be compensated by
exponential increases in stop masses, which in turn involves exponential
increases in $M_3 (0)$, and hence from Eqs.\ref{MZ},
\ref{Deltamu} exponential increases in fine-tuning.  
The exponential sensitivity of fine-tuning to the Higgs mass 
(for a fixed $\tan \beta$) was observed previously numerically \cite{price}.

It is clear that 
low values of $\tan \beta$ are associated with large
fine-tuning for two quite distinct reasons.
The first reason we saw from Eqs.\ref{MZ}, \ref{Deltamu}, 
where the $M_3(0)$ coefficient grows alarmingly
as $\tan \beta$ approaches unity.
The second reason is that we have just seen that $M_3(0)$ itself
(and to a lesser extent $A_t(0), m_{Q3}(0), m_{U3}(0)$)
must grow exponentially if the Higgs mass is to remain acceptable
as $\tan \beta$ is reduced. 
Any reduction in $M_3(0)$, such as that proposed in the previous
section, to alleviate
the fine-tuning problem, will clearly lead to a reduction in stop
masses, and hence the contribution from radiative corrections to the
Higgs mass. Of course such a reduction in the Higgs mass is easily
compensated by increasing $\tan \beta$ slightly which will
readily yield a compensating tree-level contribution. 
Thus, provided $\tan \beta$ is not too low, fine-tuning as a result
of the non-observation of Higgs at LEP may also be avoided.
If however one insists upon having very low values of $\tan \beta$ close to
unity, then correlations between $M_3(0)$ and parameters
such as $m_{H_U}(0)$
may allow large values of these parameters, and hence acceptable Higgs
masses, without fine-tuning. However even in this case there will
be a lower limit on $\tan \beta$ once radiative correction parameters
such as $\Delta \rho$ are taken into account \cite{Higgs}.

\section{Numerical Estimates}
In order to see the effect of lowering $M_3(0)$, and of correlations, 
as a function of $\tan \beta$, it is necessary 
to make some quantitative estimates of the fine-tuning parameters
for different input parameters and to examine the corresponding 
physical spectra.

\begin{figure}
\begin{center}
\leavevmode   
\hbox{
\epsfysize=5.0in
\epsffile{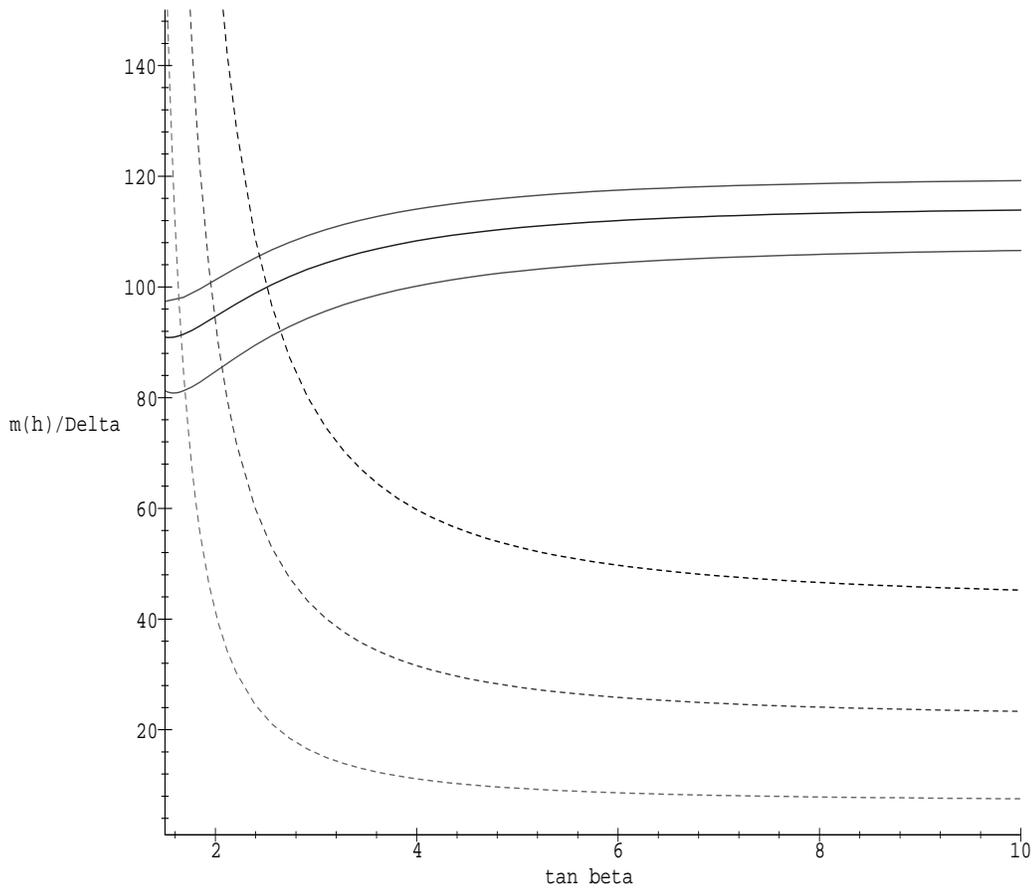}}
\end{center}
\caption
{\footnotesize The Higgs mass $m_h$ (in GeV) (solid) 
and fine-tuning parameter 
$\Delta_{\mu(0)}$ (dashes) as a function of
$\tan \beta$. 
The upper to lower curves
correspond to $M_3(0)=200,150,100$ GeV.
Apart from the gluino mass we use universal parameters 
$m_0=100$ GeV, $M_{1,2}(0)=200$ GeV in each case.
Note the large decrease in the
measure of fine-tuning as $M_3(0)$ decreases.}
\label{tanb1}
\end{figure}

Figure \ref{tanb1} shows how the Higgs mass $m_h$ and fine-tuning
parameter
$\Delta_{\mu(0)}$ vary as a function of $\tan \beta$
for $M_3(0)=200,150,100$ GeV with the other soft masses held at
universal values $m_0=100$ GeV, $M_{1,2}(0)=200$ GeV in each case.
The Higgs mass has the full one-loop radiative corrections
included. As $M_3(0)$ is reduced the fine-tuning 
parameter $\Delta_{\mu(0)}$ drops like
a stone, as does $\Delta_{M_3(0)}$ which will be discussed
separately. From the point of view of fine-tuning the $M_3(0)=100$ GeV 
case is clearly preferred with $\tan \beta \approx 2.5-10$ giving
$\Delta_{\mu(0)} \approx 20-10$ and 
$m_h \approx 90-107$ GeV.
The lower value of $M_3(0)$ leads to lower
squark and gluino masses, and lighter stops which cause the
Higgs mass to be reduced. Another direct effect of the reduction
of fine-tuning is the decrease in the value of $\mu (M_Z)$
which causes chargino and neutralino masses 
to fall and become more Higgsino-like, especially for larger $\tan \beta$.
The lighter chargino (of mass $m_{\tilde{\chi}_1^\pm} \approx 100-88$ GeV 
for $\tan \beta \approx 2.5-10$)
and neutralino may be made heavier by
increasing $M_{1,2}$, without any additional fine-tuning expense.
In the limit $\mu (M_Z) \ll M_2(M_Z)$, where the lighter chargino mass
is essentially just $\mu (M_Z)$, LEP limits on the lighter chargino
translate directly into limits on $\mu (M_Z)$. Since the fine-tuning
parameter $\Delta_{\mu (0)}$ in Eq.\ref{Deltamu} has a strong 
$\mu$ dependence the LEP limits on chargino masses will
always have some effect on fine-tuning, but in practice 
the lighter chargino mass may be increased
by increasing $M_2(0)$ without incurring any additional fine-tuning expense.
Finally we note that
$\alpha_3(M_Z)$ decreases by about 5\% over the range $M_3(0)=200-100$
GeV, due to the decreasing gluino and squark mass thresholds.
Including the gluino mass one-loop radiative corrections
\cite{gluinorc} we find $m_{\tilde{g}}\approx 284$ GeV independently
of $\tan \beta$ for $M_3(0)=100$ GeV.

\begin{figure}
\begin{center}
\leavevmode   
\hbox{
\epsfysize=5.0in
\epsffile{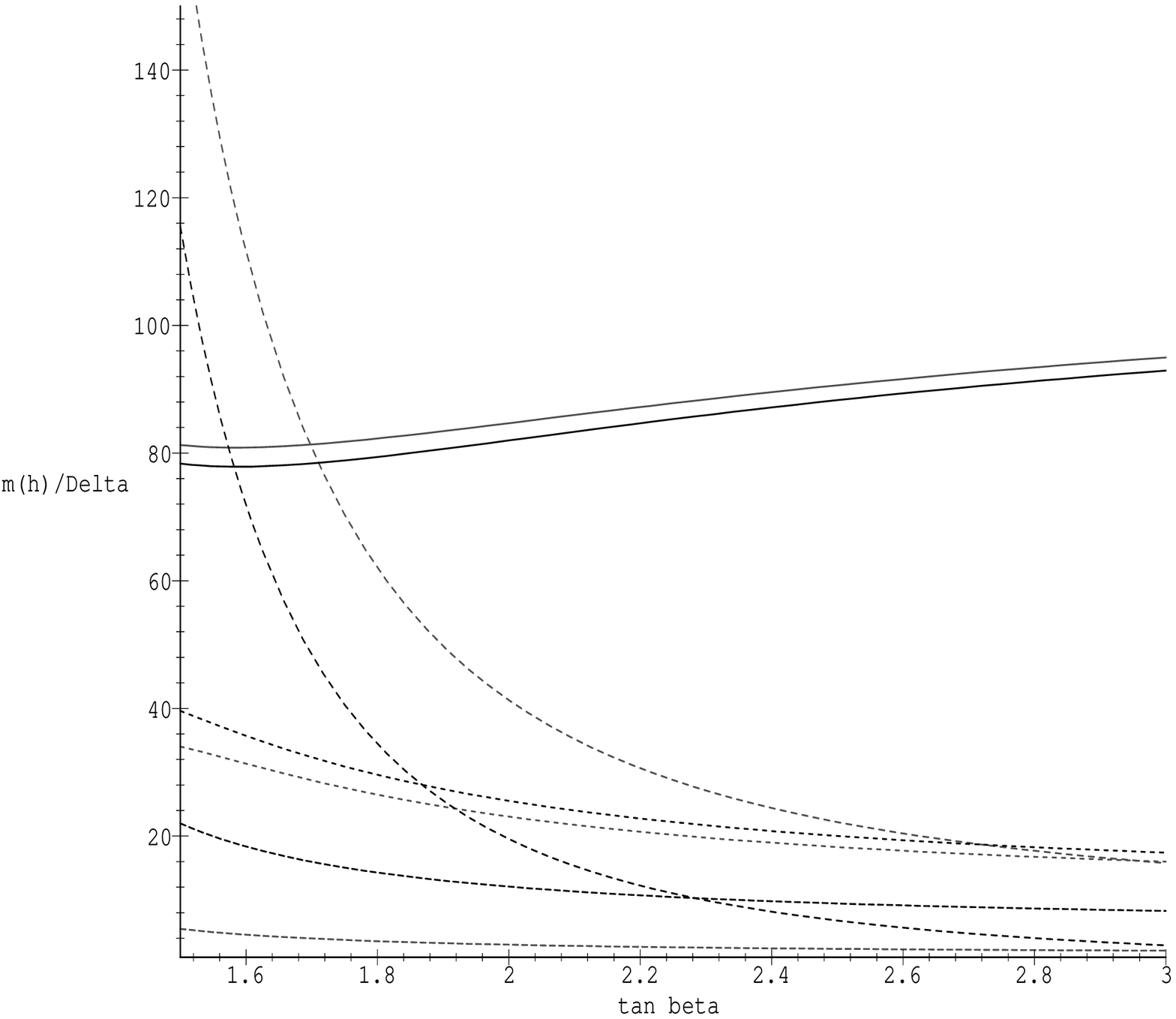}}
\end{center}
\caption
{\footnotesize 
The Higgs mass $m_h$ (in GeV) (solid) 
and fine-tuning parameters (dashes)
$\Delta_{\mu(0)}$, 
$\Delta_{M_3(0)}$,
$\Delta_{m_{H_U}^2(0)}$ 
as a function of
$\tan \beta$, for two parameter sets.
The first parameter set corresponds to
$M_3(0)=m_0=100$ GeV, $M_{1,2}(0)=200$ GeV 
and the second parameter set has
all parameters unchanged apart from a larger
second Higgs mass parameter $m_{H_U}(0)=200$ GeV.
The $m_h$, $\Delta_{\mu(0)}$ curves are obviously identified by comparison to
Figure \ref{tanb1}. In fact the uppermost values of
$m_h$, $\Delta_{\mu(0)}$ in this Figure
correspond exactly to the lowermost values in Figure \ref{tanb1}
which also uses the first parameter set here; 
note the change of scale for $\tan \beta$.  
The four remaining dashed lines correspond to
$\Delta_{M_3(0)}$ (higher pair) and $\Delta_{m_{H_U}^2(0)}$
(lower pair), where now $m_{H_U}(0)=200$ GeV gives the upper
curve of each pair.
Observe here that by increasing $m_{H_U}(0)$ relative to $M_3(0)$ there is
an additional decrease in the fine-tuning parameter 
$\Delta_{\mu(0)}$ which is
particularly relevant for small $\tan \beta$ where
it may be needed.}
\label{tanb2}
\end{figure}

It is clear from Figure \ref{tanb1} that the idea of reducing $M_3(0)$,
whilst leading to substantial reductions in fine-tuning for 
$\tan \beta \gsim 2.5$ is not by itself able to lead to
a natural theory for lower values of $\tan \beta$.
In order to reduce fine-tuning further in the
low $\tan \beta$ region we would ideally like
to reduce $M_3(0)$ significantly 
below 100 GeV, but we are effectively prevented from doing so
because this would result in the Higgs mass becoming too light.
The reduction in Higgs mass cannot be countered by
increasing the soft masses
$m_Q(0), m_U(0)$ which only leads to relatively modest
increases in Higgs mass, which is dominated by $M_3(0)$
according to Eqs.\ref{rc} and \ref{det},
and only results in additional fine-tuning in
$\Delta_{\mu(0)}$ according to Eq.\ref{Deltamu}.
Therefore in order to reduce fine-tuning further we turn
to the idea mentioned in section 2 which is to 
have a correlation or relation
$m_{H_U}(0)\approx 2M_3(0)$. 
If it is a genuine correlation then the fine-tuning is
automatically reduced by an amount which depends on
how accurately the two quantities cancel.
However if it is simply a relation then
$m_{H_U}(0)$ cannot be increased abitrarily because it will
have its own fine-tuning parameter which should not be too large.
Therefore it becomes a quantitative question as to how much
fine-tuning can genuinely be reduced in this case
before the fine-tuning parameter $\Delta_{m_{H_U}^2(0)}$ becomes
so large that additional correlations may be
relevant. Figure \ref{tanb2} addresses this question.

In Figure \ref{tanb2} we plot 
the Higgs mass $m_h$
and fine-tuning parameters 
$\Delta_{\mu(0)}$,
$\Delta_{M_3(0)}$,
$\Delta_{m_{H_U}^2(0)}$
for two parameter sets, concentrating on the low
$\tan \beta$ region.
The first parameter set is already familiar from
Figure \ref{tanb1} and corresponds to 
the smallest fine-tuning shown there
($M_3(0)=m_0=100$ GeV, $M_{1,2}(0)=200$ GeV).
In Figure \ref{tanb2} this corresponds to the upper
Higgs mass and $\Delta_{\mu(0)}$ curves and the lower
$\Delta_{M_3(0)}$ and $\Delta_{m_{H_U}^2(0)}$ curves.
For the first parameter set,
it is clearly seen that
$\Delta_{\mu(0)}$ used in Figure \ref{tanb1} is the dominant source
of fine-tuning up to $\tan \beta =3$ beyond which
$\Delta_{M_3(0)}$ becomes more important.
For $\tan \beta \lsim 2.5$, $\Delta_{\mu(0)}$ grows sharply.
By contrast $\Delta_{m_{H_U}^2(0)}$ is always very
small and corresponds to the lowest line running along the bottom
of the figure.  

The second parameter set in Figure \ref{tanb2} only differs by having
$m_{H_U}(0)=2M_3(0)=200$ GeV, motivated by the
qualitative arguments presented in section 2. This results
in only a 3 GeV decrease in the physical Higgs mass corresponding
to the lower solid curve. However it also results
in a large reduction in $\Delta_{\mu(0)}$  
corresponding to the lower large-dashed curve,
which permits smaller $\tan \beta$ to be reached for a given
amount of fine-tuning. The difference between the
$\Delta_{\mu(0)}$ curves
in Figure \ref{tanb2} (the reduction in fine-tuning )
is about the same as between pairs of $\Delta_{\mu(0)}$ curves in Figure
\ref{tanb1} where the reduction in Higgs mass is much greater.
Furthermore $\Delta_{m_{H_U}^2(0)}$,
although larger than before, is only about a half of 
$\Delta_{M_3(0)}$, which shows that at least for
$m_{H_U}(0)=200$ GeV there is no additional price to pay there.
We conclude that for small $\tan \beta$
there is a significant
reduction in fine-tuning for the case $m_{H_U}(0)\approx 2M_3(0)$
as compared to $m_{H_U}(0)\approx M_3(0)$, with only a relatively
small reduction in $m_h$. \footnote{ In Figure \ref{tanb2} for very low 
$\tan \beta$ the Higgs mass is around 80 GeV, 
but clearly it may be raised by increasing
$M_3(0)$ (which we have taken to be the round figure of 100 GeV 
for convenience only) without changing any of our qualitative conclusions.}
For larger $\tan \beta $ where fine-tuning is
controlled by $\Delta_{M_3(0)}$, this mechanism does not reduce
fine-tuning, but in that region fine-tuning is relatively small
in any case. We note that if one attempts to increase
$m_{H_U}(0)$ much further then
$\Delta_{m_{H_U}^2(0)}$ will dominate
the fine-tuning. 
\footnote{According to Eq.\ref{highparameters}
we have regarded $m_{H_U}^2(0)$ rather than $m_{H_U}(0)$ 
as an input parameter because in principle $m_{H_U}^2(0)$ can be
negative. This means that the fine-tuning parameter 
$\Delta_{m_{H_U}^2(0)}$ is only half as large as $\Delta_{m_{H_U}(0)}$
which is already
about the same size as $\Delta_{M_3(0)}$ in Figure \ref{tanb2}
for $m_{H_U}(0)=200$ GeV. }
Clearly then there is always a limit to how
low $\tan \beta$ can become before fine-tuning grows uncontrollably,
even using both low $M_3(0)$ and $m_{H_U}(0)\approx 2M_3(0)$
at the same time. If $\tan \beta$ is in fact very near unity, then
a simple relation is insufficient, and we need to appeal
to a genuine correlation between the input parameters.

Finally, on a phenomenological note, we remark that
low $\tan \beta$ is often accompanied by a light stop squark.
In this case, if the gluino is 
lighter than all the other squarks (achieved for 
example by $M_3(0)=100$ GeV, $m_0=200$ GeV)
the dominant decay of
the gluino will be $\tilde{g} \rightarrow \tilde{t}_1+t$ which
may be observed at the Tevatron, and which may be the source of many 
tops there \cite{gluino}.

\section{Conclusions}

By analyzing EWSB without universality assumptions we are led to
conclusions quite different from the usual ones.  Our first
observation is that the gaugino masses
$M_1$ and $M_2$ are hardly constrained by the EWSB conditions, 
and also the $\mu$ parameter is not as severely constrained as $M_3$,
so the masses of charginos and neutralinos can be larger without affecting
fine-tuning. {\em Therefore the absence of charginos at LEP
does not imply large fine-tuning.} 
Note that this conclusion is in agreement with that of Wright \cite{Wright}
who sets fine-tuning limits on charginos and neutralinos
of about 160 GeV.
The question of the Higgs at LEP is much more interesting and subtle. 
For $\tan \beta \gsim 2.5$ and $M_3(0) \approx 100 $ GeV it is clear
from Figure \ref{tanb1} that the present LEP limits on the Higgs mass
involve relatively small fine-tuning, although 
more stringent future Higgs limits will require 
larger $\tan \beta$, or larger $M_3$ which would 
result in increased fine-tuning.
Smaller values of $\tan \beta$ can be achieved
without increasing fine-tuning by having $m_{H_U}(0) \approx 2M_3(0)$, in
addition to a lighter gluino mass, as illustrated in Figure \ref{tanb2}.

Of course fine-tuning is not a well defined concept so 
firm quantitative conclusions are difficult to draw. However
regardless of one's measure of fine-tuning the main
qualitative constraint is
on the $SU(3)$ gaugino mass $M_3$, and therefore on the
gluino mass, which cannot get too large. We have identified 
two new ways to significantly reduce 
fine-tuning: $M_{\tilde g} \lsim 350$ GeV, 
and/or there is some correlation between $m_{H_U}$ and $M_3$, 
approximately that $m_{H_U}(0) \approx 2M_3(0)$, 
perhaps with additional contributions from squark soft
masses or $A_t$ or $\mu$. A strict correlation might not
be necessary since even a simple relation between these two
input parameters, which are regarded as independent, may reduce
fine-tuning as shown in Figure \ref{tanb2}. Another possible correlation
is between $\mu$ and $M_3$ if $\mu$ is large.
It is very interesting that a lighter gluino pushes down the predicted
$\alpha_3(M_Z)$ towards the experimental value,
as discussed in the introduction.

In this paper
we have made no effort to fit or optimize fine-tuning parameters, because
we do not believe any particular measure or value is much to be preferred.
But we do believe that the general notion of a fine-tuning constraint is
real and important, and we think it is telling us significant information
about the soft-breaking masses.  Most likely $M_3$ is smaller than its
universality value relative to $M_1$ and $M_2$, 
and in addition perhaps $M_3$ and
$m_{H_U}$ are related, particularly if $\tan \beta$ is small.
Both ways we learn
something about unification scale physics from collider data --- either
gaugino mass universality is violated (perhaps by as much as a factor of
two), or there is a relation among soft-breaking parameters, or both. 
It will be possible to distinguish between these at the
Fermilab upgraded collider.

\begin{center}
{\bf Acknowledgements}
\end{center}
SFK would like to thank Fermilab and University of Michigan
Theory Groups for hospitality extended,
and Marcela Carena for discussions.
GLK appreciates discussions with Stefan Pokorski and Lisa Everett.

\end{document}